\documentclass[conference]{IEEEtran}
\IEEEoverridecommandlockouts
\usepackage{cite}
\usepackage{amsmath,amssymb,amsfonts}
\usepackage{algorithmic}
\usepackage{graphicx}
\usepackage{float}
\usepackage{textcomp}
\usepackage{xcolor}
\usepackage{colortbl}
\usepackage{url}
\usepackage{multirow}
\usepackage{graphicx}
\usepackage[keeplastbox]{flushend}
\usepackage{balance}

\def\BibTeX{{\rm B\kern-.05em{\sc i\kern-.025em b}\kern-.08em
    \textit{T}\kern-.1667em\lower.7ex\hbox{E}\kern-.125emX}}

\begin{document}

\title{Link Budget Analysis for Free-Space Optical Satellite Networks}

\author{\IEEEauthorblockN{Jintao Liang\textsuperscript{$1$}, Aizaz U. Chaudhry\textsuperscript{$2$}, Eylem Erdogan\textsuperscript{$3$}, and Halim Yanikomeroglu\textsuperscript{$2$}}
\IEEEauthorblockA{\textsuperscript{$1$}\textsuperscript{$2$}Department of Systems and Computer Engineering, Carleton University, Ottawa, Canada -- K1S 5B6 \\
\textsuperscript{$3$}Department of Electrical and Electronics Engineering, Istanbul Medeniyet University, Uskudar, Istanbul, Turkey – 34700 \\
\textsuperscript{$1$}jintaoliang@cmail.carleton.ca, \textsuperscript{$2$}\{auhchaud, halim\}@sce.carleton.ca, \textsuperscript{$3$}eylem.erdogan@medeniyet.edu.tr}
}

\maketitle

\begin{abstract}
Free-space optical satellite networks (FSOSNs) will employ free-space optical links between satellites and between satellites and ground stations, and the link budget for optical inter-satellite links and optical uplink/downlink is analyzed in this paper. The satellites in these FSOSNs will have limited energy and thereby limited power, and we investigate the effect of link distance and link margin on optical inter-satellite link transmission power, and the effect of slant distance, elevation angle, and link margin on optical uplink/downlink transmission power. We model these optical links and compute the results for various parameters. We observe that the transmission power increases when the link distance increases for inter-satellite and uplink/downlink communications, while the transmission power decreases when the elevation angle increases for uplink/downlink transmission. We also observe an inverse relationship between link margin and link distance. Furthermore, we highlight some practical insights and design guidelines gained from this analysis.  \\
\end{abstract}

\section{Introduction}
Free-space optical communication is receiving more and more attention these days as it is a promising and rapidly developing technology for wireless communication between satellites due to its larger link bandwidth, license free spectrum, higher link data rate, better security, smaller antenna size, lower terminal mass, and lower terminal power consumption compared to radio frequency-based satellite communication \cite{b1}. The free-space optical satellites networks (FSOSNs) based on upcoming low Earth orbit (LEO) satellite constellations, such as SpaceX’s Starlink \cite{b3},  and Telesat’s Lightspeed \cite{b5}, are expected to employ optical or laser inter-satellite links while optical or laser uplink and downlink communications are envisioned for future FSOSNs\cite{b6}, \cite{b7}. 

A satellite harvests solar energy via its solar panels and stores it in its battery. The lifetime of a satellite’s battery is determined by the number of charging and recharging cycles, and a satellite’s working lifetime is equal to its battery’s lifetime. If the transmission power of a link between satellites or between a satellite and a ground station is high, the satellite needs more energy from the battery, which results in more frequent discharging/charging and leads to shorter battery lifetime and thereby shorter satellite lifetime. Once the battery is depleted, the satellite will lose power, become inoperative, and need to be replaced and de-orbited. In this way, the analysis and design of link budget for FSOSNs is important.

In this work, we analyze the link budget for communication over optical inter-satellite links and optical uplink and downlink communications. Since satellites are orbiting around the Earth in space, the propagation medium for inter-satellite communication is vacuum of space, and the optical beams propagate without attenuation and fading due to the propagation medium. For uplink and downlink, the optical beams must go through the atmosphere, which causes multiple attenuation and fading due to scattering and scintillation, and the required transmission power should be adjusted accordingly to compensate for the loss caused by the atmosphere. 

In the link budget analysis of FSOSNs, we first vary the link distance and set the required received power as -35.5 dBm, data rate as 10 Gbps, and bit error rate as $10^{-12}$ \cite{b8}. We observe that the link transmission power increases with the increase in link distance. Then, we vary the elevation angle for uplink/downlink and fix the altitude of the satellite. We find out that the atmospheric attenuation increases when the elevation angle decreases. We also investigate the relationship between link margin and link distance for both optical inter-satellite link and optical uplink/downlink, and the results show an inverse relationship between the two. Furthermore, some practical insights and design guidelines that emerge from this anaylsis are discussed. To the best of our knowledge, for the first time in literature we investigate transmission power in terms of link distance, link margin, and elevation angle for both optical inter-satellite link and optical uplink/downlink.

The rest of the paper is organized as follows. The related work is discussed in Section II. Section III presents the system model, including the link budget modelling for optical inter-satellite link and optical uplink/downlink. Section IV provides the results and analysis of various factors affecting link transmission power. Section V discusses practical insights and design guidelines. Conclusions and future work are summarized in Section VI.

\section{Related Work}
In the literature, there are various studies that consider link budget analysis for optical satellite communications as can be seen in \cite{b8}--\!\!\cite{b12} and the references therein. Different from the current literature, we investigate both optical inter-satellite link between LEO satellites and optical uplink/downlink between LEO satellites and ground stations that can establish FSOSNs in space. In \cite{b8},  the authors study the bit error rate vs. received power for optical inter-satellite links, while we focus on examining the factors affecting link transmission power for optical inter-satellite link and optical uplink/downlink in LEO FSOSNs. In \cite{b9}, the authors investigate the optical inter-satellite link with data rate of 10 Gbps and the relationship between link margin and link distance, while in this work we also study the link margin and slant distance for optical uplink/downlink. 

In \cite{b10}, the authors mention the link budget model for optical inter-satellite link and simulate links using QPSK modulation to find relationship between link distance and data rate, while in this work we assume on-off keying (OOK) and simulate the optical links with more practical parameters. In \cite{b11}, the authors give the model for optical link budget and link margin, but their analysis is based on simulation of LEO-to-GEO and GEO-to-ground optical links, while we focus on LEO-to-LEO and LEO-to-ground optical links. In \cite{b12}, Mie scattering, and geometrical scattering are considered in atmospheric attenuation for optical uplink/downlink, while in this work we also investigate the effect of slant distance and elevation angle on atmospheric attenuation for optical uplink/downlink.

\section{System Model}
This section introduces the system model for optical links in FSOSNs, which includes optical inter-satellite link model and optical uplink/downlink model as well as link margin model. 

The geometrical expression of the parameters in the inter-satellite link and uplink/downlink can be found in Fig. 1. In this figure, the blue circular bound around the surface of the Earth refers to the troposphere layer of the atmosphere with height $\textit{h}_A$ km and \textit{O} refers to the center of the Earth. The ground station is located at $\textit{h}_E$ km above the mean sea level, and the elevation angle (i.e., the angle between the tangential line to the surface of the Earth, shown as dashed line in Fig. 1, and the link between the ground station and the satellite Sat A) is $\textit{$\theta$}_E$ degrees. $\textit{R}_E$ is the radius of the Earth and is considered as 6,378.1 km, and $\textit{h}_S$ is the altitude of the satellites Sat A and Sat B. The distance from \textit{O} to Sat A is $\textit{R}_E$ $+$ $\textit{h}_S$ and the distance from $\textit{O}$ to ground station is $\textit{R}_E$ $+$ $\textit{h}_E$, $\textit{d}_A$ represents the distance that the optical laser beam propagates through the troposphere layer of the atmosphere, and $\textit{d}_{GS}$ is the slant distance for the uplink and downlink between satellite Sat A and ground station. $\textit{d}_{SS}$ is the distance between the two satellites Sat A and Sat B.

\begin{figure}[htbp]
\centerline{\includegraphics[width=3.4 in]{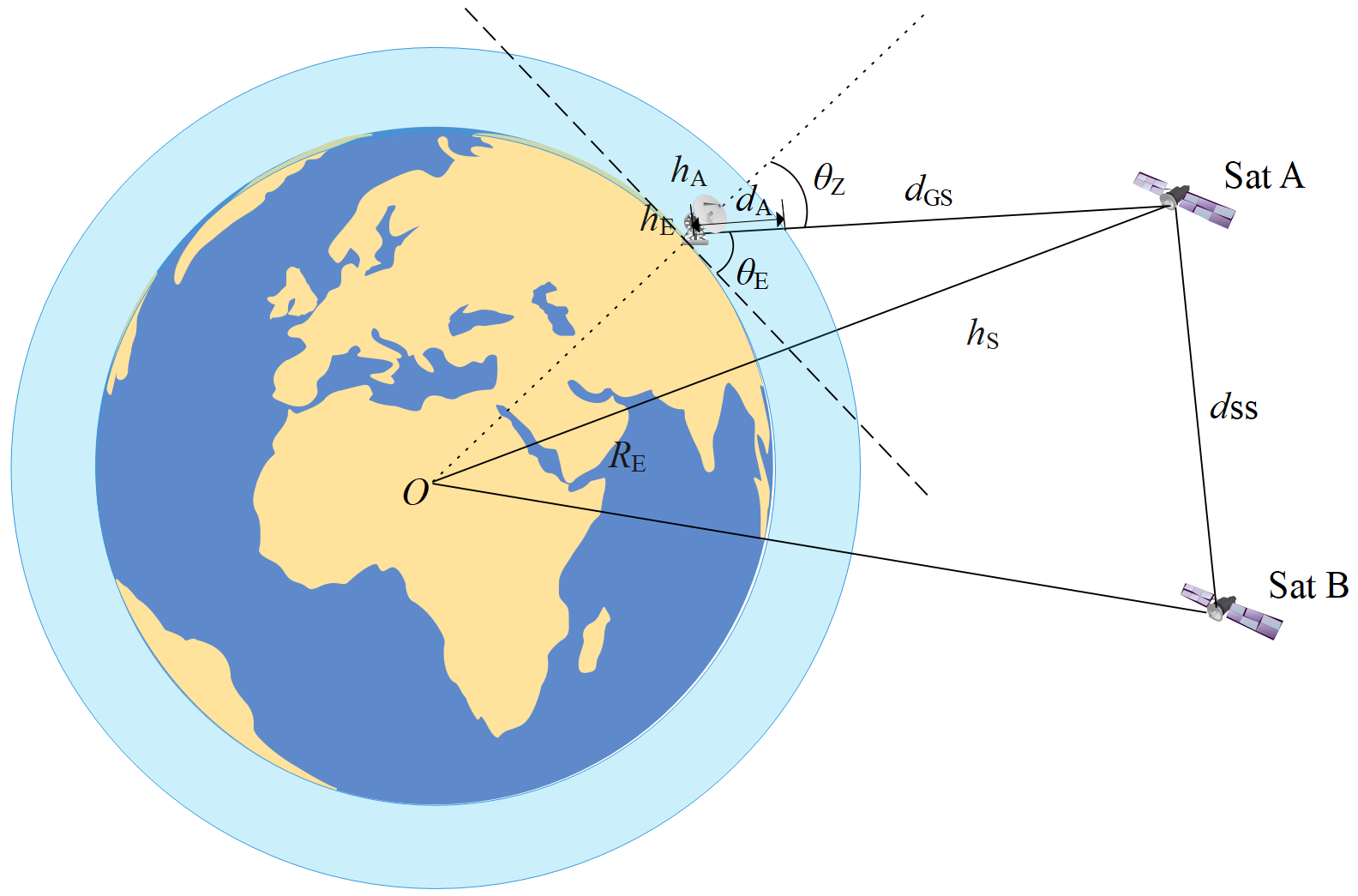}}
\caption{Geometrical representation of parameters for inter-satellite link and uplink/downlink optical communication.}
\label{fig}
\end{figure}

\subsection{Optical Inter-Satellite Link Model}

For FSOSNs, the optical links between transmitters and receivers can be classified as uplink, inter-satellite link and downlink. An optical inter-satellite link is the link between two satellites, and the propagation medium is the vacuum of space since these links exist between satellites located in space. For an optical inter-satellite link, the received power is given as 
\begin{equation}
\label{eq_1}
\textit{P}_R = \textit{P}_T\textit{$\eta$}_T\textit{$\eta$}_R\textit{G}_T\textit{G}_R\textit{L}_T\textit{L}_R\textit{L}_{PS} \text{,}
\end{equation}
where $\textit{P}_R$ is the received power in Watts, $\textit{P}_T$ is the transmitted power in Watts, $\textit{$\eta$}_T$ is the optics efficiency of the transmitter, $\textit{$\eta$}_R$ is the optics efficiency of the receiver, $\textit{G}_T$ is the transmitter gain, $\textit{G}_R$ is the receiver gain, $\textit{L}_T$ is the transmitter pointing loss, $\textit{L}_R$ is the receiver pointing loss, and $\textit{L}_{PS}$ is the free-space path loss for the optical link between satellites \cite{b13}. The transmitter gain $\textit{G}_T$ in (1) is expressed as $\textit{G}_T = 16/({\textit{$\Theta$}_T})^{2}$, where $\textit{$\Theta$}_T$ is the full transmitting divergence angle in radians \cite{b14}; the receiver gain $\textit{G}_R$ is expressed as $\textit{G}_R = (\textit{D}_R\textit{$\pi$}/\textit{$\lambda$})^2$, where $\textit{D}_R$ is the receiver’s telescope diameter in mm \cite{b14}; the transmitter pointing loss $\textit{L}_T$ is given as $\textit{L}_T = \exp{(-\textit{G}_T(\textit{$\theta$}_T)^{2})}$, where $\textit{$\theta$}_T$ is the transmitter pointing error in radians \cite{b13}; the receiver pointing loss $\textit{L}_R$ is written as $\textit{L}_R = \exp{(-\textit{G}_R(\textit{$\theta$}_R)^{2})}$, where $\textit{$\theta$}_R$ is the receiver pointing error in radians \cite{b13}; and the free-space path loss $\textit{L}_{PS}$ is given as
\begin{equation}
\label{eq_2}
\textit{L}_{PS} = (\textit{$\lambda$}/4\textit{$\pi$}\textit{d}_{SS})^2\text{,} 
\end{equation}  
where $\textit{$\lambda$}$ is the operating wavelength in nm, and $\textit{d}_{SS}$ is the distance between satellites in km \cite{b13}.

\subsection{Optical Uplink/Downlink Model}
Optical uplink and downlink communications between satellites and ground stations experience attenuations because of the atmosphere. Scattering is generally defined as the redirection of beam energy by particles present along the beam propagation path. In this work, we consider Mie scattering and geometrical scattering to model atmospheric attenuation as these two are the primary sources of beam scattering and thereby beam fading in the atmosphere. 

To model optical uplink and downlink, atmospheric attenuation must be considered and the received power is given as
\begin{equation}
\label{eq_3}
\textit{P}_R = \textit{P}_T\textit{$\eta$}_T\textit{$\eta$}_R\textit{G}_T\textit{G}_R\textit{L}_T\textit{L}_R\textit{L}_A\textit{L}_{PG}\text{,}
\end{equation}
where $\textit{L}_A$ is the atmospheric attenuation loss, $\textit{L}_{PG}$ is the free-space path loss for links between ground stations and satellites \cite{b11} and other parameters are like the optical inter-satellite link model in (1). The slant distance (i.e., the distance between a ground station and a satellite) for uplink/downlink $\textit{d}_{GS}$ is given as
\begin{equation}
\label{eq_4}
\textit{d}_{GS} = \textit{R}({\sqrt{((\textit{R}+ \textit{H})/\textit{R})^2-(\cos({\textit{$\theta$}_E}))^{2}}-\sin({\textit{$\theta$}_E}}))\text{,}
\end{equation} 
where $\textit{R}$ $=$ $\textit{R}_E$ $+$ $\textit{h}_E$ and $\textit{H}$ $=$ $\textit{h}_S$ $-$ $\textit{h}_E$ \cite{b15}. The free-space path loss $\textit{L}_{PG}$ can be expressed based on slant distance as
\begin{equation}
\label{eq_5}
\textit{L}_{PG} = (\textit{$\lambda$}/4\textit{$\pi$}\textit{d}_{GS})^2\text{.} 
\end{equation}
We consider the altitude of the ground station in (4) to model practical scenarios as in real cases the ground stations are mostly located at high places on the surface of the Earth.
\vspace*{2pt}

\subsubsection{Atmospheric Attenuation due to Mie Scattering}
\hspace*{\fill}\\
\indent
Mie scattering occurs when the diameter of atmospheric particles is equal to or greater than the wavelength of the optical beam. It mainly occurs in the lower part of the atmosphere where larger particles are more abundant, and it is primarily caused by microscopic particles of water. The following expression, which can precisely model the Mie scattering effect, is appropriate for ground stations located at altitudes between 0 and 5 km above the mean sea level:
\begin{equation}
\label{eq_6}
\textit{$\rho$} = \textit{a}(\textit{h}_E)^{3}+\textit{b}(\textit{h}_E)^{2}+\textit{c}\textit{h}_E+\textit{d}\text{,}
\end{equation}  
where $\textit{$\rho$}$ denotes the extinction ratio, $\textit{h}_E$ is the height of the ground station above the mean sea level in km, and $\textit{a}$, $\textit{b}$, $\textit{c}$ and $\textit{d}$ are the wavelength dependent empirical coefficients, which can be expressed as $\textit{a} = -0.000545\textit{$\lambda$}^2+0.002\textit{$\lambda$}-0.0038$, $\textit{b} = 0.00628\textit{$\lambda$}^2-0.0232\textit{$\lambda$}+0.00439$, $\textit{c} = -0.028\textit{$\lambda$}^2+0.101\textit{$\lambda$}-0.18$, $\textit{d} = -0.228\textit{$\lambda$}^3+0.922\textit{$\lambda$}^2-1.26\textit{$\lambda$}+0.719$, and the atmospheric attenuation due to Mie scattering can be expressed as
\begin{equation}
\label{eq_7}
\textit{I}_m = \exp{(-\textit{$\rho$}/\sin({\textit{$\theta$}_E}))}\text{,}   
\end{equation} 
where $\textit{$\theta$}_E$ is the elevation angle of the ground station in degrees \cite{b16}.
\vspace*{2pt}

\subsubsection{Atmospheric Attenuation due to Geometrical Scattering}
\hspace*{\fill}
Geometrical scattering is used to model the attenuation due to atmosphere that is close to the surface of the Earth and is caused by fog or dense clouds. In this model, the following expression shows the effect of geometrical scattering:
\begin{equation}
\label{eq_8}
\textit{V} = 1.002/(\textit{L}_W\textit{N})^{0.6473}\text{,}   
\end{equation} 
where $\textit{V}$ is the visibility in km, $\textit{L}_W$ is the liquid water content in $g/m^{-3}$ and $\textit{N}$ is the cloud number concentration in $cm^{-3}$. The attenuation coefficient $\textit{$\theta$}_A$ can be expressed as
\begin{equation}
\label{eq_9}
\textit{$\theta$}_A = (3.91/\textit{V}) (\textit{$\lambda$}/550)^{-\textit{$\varphi$}}\text{,}    
\end{equation}
where $\textit{$\varphi$}$ is the particle size related coefficient given according to Kim’s model \cite{b17}. The Beer-Lambert law is given as $\textit{I}(\textit{z})= \exp{(-\textit{$\mu$}\textit{z})}$, where $\textit{$\mu$}$ is the attenuation coefficient that depends on wavelength and $\textit{z}$ is the distance of the transmission path \cite{b18}. For geometrical scattering, the atmospheric attenuation can be expressed using the Beer-Lambert law as
\begin{equation}
\label{eq_10}
\textit{I}_g= \exp{(-\textit{$\theta$}_A\textit{d}_A)}\text{,}    
\end{equation} 
where $\textit{d}_A$ is the distance of the optical beam through the troposphere layer of the atmosphere over which it encounters geometrical scattering, and it can be expressed based on the zenith angle (i.e., the angle between the perpendicular to the surface of the Earth, shown as dotted line in Fig. 1, and the link between the ground station and the satellite Sat A) $\textit{$\theta$}_Z$ as $\textit{d}_A= (\textit{h}_A-\textit{h}_E)\sec({\textit{$\theta$}_Z})$ \cite{b19}, and it can also be calculated using the elevation angle $\textit{$\theta$}_E$ as $\textit{d}_A= (\textit{h}_A-\textit{h}_E)\csc({\textit{$\theta$}_E})$, where $\textit{h}_A$ is the height of the troposphere layer of atmosphere in km, $\textit{h}_E$ is the altitude of the ground station in km, and $\textit{$\theta$}_E$ = 90º $-$ $\textit{$\theta$}_Z$. 

The atmospheric attenuation loss considering both Mie scattering and geometrical scattering \cite{b12} can then be calculated as 
\begin{equation}
\label{eq_11}
\textit{L}_A = \textit{I}_m\textit{I}_g = \exp{(-\textit{$\rho$}/\sin({\textit{$\theta$}_E}))}\exp{(-\textit{$\theta$}_A\textit{d}_A)}\text{.}    
\end{equation}

\subsection{Link Margin Model} 
The performance of an optical communication system is commonly evaluated in terms of link margin, bit error rate and etc. The link margin is defined as the ratio of the received signal power and the required signal power that is needed to achieve a specific bit error rate at a given data rate. The link margin is needed to counter unexpected losses and noises, and it should be always positive to guarantee that the received signal can be received properly. The link margin can be modelled as 
\begin{equation}
\label{eq_12}
\textit{LM} = \textit{P}_R/\textit{P}_{req}\text{,}    
\end{equation}
where $\textit{P}_R$ is the received power in mW and $\textit{P}_{req}$ is the receiver sensitivity in mW \cite{b11}. In this work, we are interested in the link transmission power. Thereby, we have to find the received power, which can be expressed based on the link margin as $\textit{P}_R = \textit{LM}$×$\textit{P}_{req}$. 

We first investigate the impact of link distance and elevation angle on link transmission power with fixed link margin, and then study the relationship between link margin and link distance for fixed transmission power.

\section{Results}
In this section, we present the numerical results obtained for link transmission power based on link distance, link margin, and elevation angle to analyze the performance of optical uplink/downlink and optical inter-satellite link. Furthermore, we investigate the relationship between link margin and link distance for these links. 

\subsection{Transmission Power vs. Link Distance for Optical Inter-Satellite Link}
We consider the optical link model parameters summarized in Table 1 to evaluate the optical inter-satellite links and uplink/downlink. These parameters are used in practical and realized optical satellite communication systems. We consider OOK as the optical link’s modulation scheme. We use curve fitting technique to find the receiver sensitivity as -35.5 dBm for OOK modulation with 10 Gbps data rate and $10^{-12}$ bit error rate from \cite{b9}. We consider $\textit{P}_{req}$ as -35.5 dBm and set the \textit{LM} as 3 dB. Based on the parameters in Table 1, we use (1) and (2) to compute the link transmission power $\textit{P}_T$ for different link distances between satellites as shown in Table 2. As the table indicates, when $\textit{d}_{ss}$ increases, $\textit{P}_T$ increases as expected.

\begin{table}
\centering
\renewcommand\thetable{1}
\caption{Optical Link Model Parameters.}
\begin{tabular}{llll}
\hline
\textbf{Parameter}                      & \textbf{Symbol}                                                                       & \textbf{Units}                                                               & \textbf{Value}          \\
\hline
{Laser wavelength \cite{b8}}                   & \textit{$\lambda$}                                                                             & {nm}                                                                           & {1550}                     \\
{Transmitter optical efficiency \cite{b10}}     & $\textit{{$\eta$}}_{{T}}${}    & {~}                                                                            & {0.8}                      \\
{Receiver optical efficiency \cite{b10}}        & $\textit{{$\eta$}}_{{R}}${}    & {~}                                                                            & {0.8}                      \\
{Data rate \cite{b9}}                          & $\textit{{R}}_{{data}}${} & {Gbps}                                                                         & {10}                       \\
{Receiver telescope diameter \cite{b8}}        & $\textit{{D}}_{{R}}${}    & {mm}                                                                           & {80}                       \\
{Transmitter pointing error \cite{b10}}         & $\textit{{$\theta$}}_{{T}}${}    & {$\mu$}{rad}{} & {1}                        \\
{Receiver pointing error \cite{b10}}            & $\textit{{$\theta$}}_{{R}}${}    & {$\mu$}{rad}{} & {1}                        \\
{Full transmitting divergence angle \cite{b8}} & $\textit{{$\Theta$}}_{{T}}${}    & {$\mu$}{rad}{} & {15}                       \\
{Receiver sensitivity \cite{b9}}                & $\textit{{P}}_{{req}}${}  & {dBm}                                                                          & {-35.5}                    \\
{Bit error rate \cite{b9}}                      & {~}                                                                                      & {~}                                                                            & {10\textsuperscript{-12}} \\
\hline
\end{tabular}
\end{table}

\begin{table}
\centering
\renewcommand\thetable{2}
\caption{Transmission Power vs. Link Distance for Optical Inter-Satellite Link.}
\arrayrulecolor{black}
\begin{tabular}{!{\color{black}\vrule}l!{\color{black}\vrule}l!{\color{black}\vrule}l!{\color{black}\vrule}l!{\color{black}\vrule}} 
\hline
$\textit{d}_{SS}$ (km) & $\textit{L}_{PS}$ (dB) & $\textit{P}_T$ (dBm) & $\textit{P}_T$ (W)                \\ 
\hline
1000              & -258.18          & 15.32             & 34.05×10\textsuperscript{-3}   \\ 
\hline
2000              & -264.20          & 21.34             & 136.20×10\textsuperscript{-3}  \\ 
\hline
3000              & -267.74          & 23.87             & 306.46×10\textsuperscript{-3}   \\ 
\hline
4000              & -270.24          & 27.36             & 544.81×10\textsuperscript{-3}  \\ 
\hline
4500              & -271.24          & 28.39             & 689.53×10\textsuperscript{-3}  \\ 
\hline
5000              & -272.18          & 29.30             & 851.27×10\textsuperscript{-3}   \\ 
\hline
5500              & -272.98          &30.13             & 1.03                          \\ 
\hline
6000              & -273.76          & 30.88             & 1.23                           \\ 
\hline
7000              & -275.10          & 32.22             & 1.67                           \\ 
\hline
8000              & -276.26          & 33.38             & 2.18                            \\ 
\hline
9000              & -277.26          & 34.41             & 2.76                          \\ 
\hline
10000             & -278.18          & 35.32             & 3.41                           \\
\hline
\end{tabular}
\arrayrulecolor{black}
\end{table}

\subsection{Transmission Power vs. Slant Distance for Optical Uplink/Downlink}
For the computation of the optical uplink/downlink transmission power, the parameters related to the atmospheric attenuation are summarized in Table 3. We use (6) and (7) to compute Mie scattering $\textit{I}_m$, which depends upon ground station altitude $\textit{h}_E$, elevation angle $\textit{$\theta$}_E$, and wavelength $\textit{$\lambda$}$. We find geometrical scattering $\textit{I}_g$ according to (8)--(10). Thereafter, we obtain the atmospheric attenuation loss $\textit{L}_A$ by using (11) and compute $\textit{P}_T$ for optical uplink/downlink as shown in Table 4 when $\textit{$\theta$}_E$ is fixed as 40º, and \textit{LM} is fixed as 3 dB. In this table, we vary the altitude of the satellite $\textit{h}_S$ and get corresponding slant distance $\textit{d}_{GS}$ using (4). With increase in $\textit{h}_S$ and thereby $\textit{d}_{GS}$, $\textit{P}_T$ increases as shown in this table.

\begin{table}
\centering
\renewcommand\thetable{3}
\caption{Atmospheric Attenuation Parameters.}
\arrayrulecolor{black}
\begin{tabular}{llll} 
\hline
\textbf{{Parameter}}                                                 & \textbf{{Symbol}}                                                                     & \textbf{{Units}}                             & \textbf{{Value}}                          \\ 
\hline
{Ground station height \cite{b12}}                                       & ${h}_{{E}}${}         & {km}                                         & {1}                                       \\
{Thin cirrus cloud concentration \cite{b17}}                               & $\textit{{L}}_{{W}}${} & {cm\textsuperscript{-3}}                     & {0.5}                                     \\
{Liquid water content \cite{b17}}                                          & \textit{{N}}                                                                          & {g/m\textsuperscript{-3}}                    & {3.128×10\textsuperscript{-4}}            \\
{Partial size coefficient \cite{b12}}                                      & \textit{{$\varphi$}}                                                                          & {~}                                          & {1.6}                                     \\
{Elevation angle \cite{b3}}{}           & $\textit{{$\theta$}}_{{E}}${} & {degree}{} & {40}{}  \\
{Troposphere layer height \cite{b19}}{} & $\textit{{h}}_{{A}}${} & {km}{}     & {20}{}  \\
\hline
\end{tabular}
\arrayrulecolor{black}
\end{table}

\begin{table*}
\centering
\renewcommand\thetable{4}
\setlength{\tabcolsep}{15pt}
\caption{Transmission Power vs. Slant Distance for Optical Uplink/Downlink.}
\arrayrulecolor{black}
\begin{tabular}{!{\color{black}\vrule}l!{\color{black}\vrule}l!{\color{black}\vrule}l!{\color{black}\vrule}l!{\color{black}\vrule}l!{\color{black}\vrule}l!{\color{black}\vrule}l!{\color{black}\vrule}l!{\color{black}\vrule}} 
\hline
$\textit{h}_S$ (km) & $\textit{d}_{GS}$ (km) & $\textit{L}_{PG}$ (dB) & $\textit{I}_m$ (dB) & $\textit{I}_g$ (dB) & $\textit{L}_A$ (dB) & $\textit{P}_T$ (dBm) & $\textit{P}_T$ (W)                \\ 
\hline
300             & 451.2              & -251.27           & -0.15           & -0.33           & -0.48           & 8.89	      &7.75×10\textsuperscript{-3}   \\ 
\hline
400             & 596.7              & -253.69          & -0.15           & -0.33           & -0.48             & 11.32             & 13.55×10\textsuperscript{-3}   \\ 
\hline
500             & 739.9              & -255.56           & -0.15           & -0.33           & -0.48             & 13.19             & 20.83×10\textsuperscript{-3}   \\ 
\hline
600             & 881.0              & -257.08          & -0.15           & -0.33           & -0.48            & 14.70             & 29.54×10\textsuperscript{-3}   \\ 
\hline
700             & 1020.1             & -258.35           & -0.15           & -0.33           & -0.48             & 15.98             & 39.60×10\textsuperscript{-3}   \\ 
\hline
800             & 1157.5             & -259.45         & -0.15           & -0.33           & -0.48             & 17.07             & 50.98  ×10\textsuperscript{-3}  \\ 
\hline
900             & 1293.2             & -260.41          & -0.15           & -0.33           & -0.48             & 18.04            & 63.64×10\textsuperscript{-3}   \\ 
\hline
1000            & 1427.4             & -261.27           & -0.15           & -0.33           & -0.48             & 18.90             & 77.54×10\textsuperscript{-3}  \\ 
\hline
1100            & 1560.2             & -262.04           & -0.15           & -0.33           & -0.48             & 19.67             & 92.63×10\textsuperscript{-3}  \\ 
\hline
1200            & 1691.7             & -262.74           & -0.15           & -0.33           & -0.48             & 20.37             & 108.90×10\textsuperscript{-3}  \\ 
\hline
1300            & 1821.9             & -263.39          & -0.15           & -0.33           & -0.48             & 21.01             & 126.31×10\textsuperscript{-3}  \\ 
\hline
1400            & 1951.0             & -263.98          & -0.15           & -0.33           & -0.48             & 21.61            & 144.85×10\textsuperscript{-3}  \\ 
\hline
1500            & 2079.0             & -264.53         & -0.15           & -0.33           & -0.48             & 22.16            & 164.48×10\textsuperscript{-3}  \\
\hline
\end{tabular}
\arrayrulecolor{black}
\end{table*}

\subsection{Transmission Power vs. Elevation Angle for Optical Uplink/Downlink}
Since both Mie and geometrical scatterings are related to elevation angle $\textit{$\theta$}_E$ between the satellite and the ground station, we further compute $\textit{P}_T$ for various elevation angles from 10º to 90º and fix the altitude of the satellite $\textit{h}_S$ as 550 km. The satellite is right above the ground station when $\textit{$\theta$}_E$ reaches 90º, and the slant distance $\textit{d}_{GS}=\textit{h}_S-\textit{h}_E$. We show the corresponding results in Table 5, which indicate that $\textit{P}_T$ decreases with increase in $\textit{$\theta$}_E$. 

\begin{table*}
\centering
\renewcommand\thetable{5}
\setlength{\tabcolsep}{11pt}
\caption{Transmission Power vs. Elevation Angle for Optical Uplink/Downlink.}
\arrayrulecolor{black}
\begin{tabular}{!{\color{black}\vrule}l!{\color{black}\vrule}l!{\color{black}\vrule}l!{\color{black}\vrule}l!{\color{black}\vrule}l!{\color{black}\vrule}l!{\color{black}\vrule}l!{\color{black}\vrule}l!{\color{black}\vrule}l!{\color{black}\vrule}} 
\hline
$\textit{$\theta$}_E$ (º) & $\textit{d}_{GS}$ (km) & $\textit{d}_A$ (km) & $\textit{L}_{PG}$ (dB) & $\textit{I}_m$ (dB) & $\textit{I}_g$ (dB) & $\textit{L}_A$ (dB) & $\textit{P}_T$ (dBm) & $\textit{P}_T$ (W)                \\ 
\hline
10              & 1813.4            & 109.4                     & -263.35           & -0.57                    & -1.22                   & -1.79                    & 22.28            & 168.96×10\textsuperscript{-3}  \\ 
\hline
20              & 1291.8            & 55.6                      & -260.40           & -0.29                   & -0.62                    & -0.91                   & 18.45            & 70.02×10\textsuperscript{-3}  \\ 
\hline
30              & 991.2             & 38.0                      & -258.10           & -0.20                    &-0.42                    & -0.62                    & 15.87             & 38.60×10\textsuperscript{-3}   \\ 
\hline
40              & 810.7             & 29.6                      & -256.35           & -0.15                    &-0.33                   &-0.48                     & 13.98             & 25.01×10\textsuperscript{-3}   \\ 
\hline
50              & 697.7             & 24.8                      & -255.05           & -0.13                    & -0.26                   & -0.41                     & 12.60             & 18.20×10\textsuperscript{-3}   \\ 
\hline
60              & 625.8             & 21.9                      & -254.11           & -0.11                    & -0.24                    & -0.36                     & 11.61             &14.48×10\textsuperscript{-3}   \\ 
\hline
70              & 581.2             & 20.2                      & -253.47           & -0.11                    & -0.22                    & -0.33                     & 10.94           & 12.41×10\textsuperscript{-3}   \\ 
\hline
80              & 556.8             & 19.3                      & -253.09           & -0.10                    & -0.21                    & -0.32                     & 10.55             & 11.35×10\textsuperscript{-3}   \\ 
\hline
90              & 549.0             & 19.0                       & -252.97           & -0.10                    & -0.21                    & -0.31                     & 10.42             & 11.02×10\textsuperscript{-3}   \\
\hline
\end{tabular}
\arrayrulecolor{black}
\end{table*}

\subsection{Transmission Power vs. Link Margin for Optical Inter-Satellite Link}
As shown in Table 6, we compute $\textit{P}_T$ for different values of $\textit{LM}$ and $\textit{d}_{SS}$. For a certain value of $\textit{d}_{SS}$ in this table, $\textit{P}_T$ increases with increase in $\textit{LM}$. We also investigate the value for $\textit{LM}$ with a given $\textit{P}_T$ based on different link distances. For optical inter-satellite link, we find the value of $\textit{LM}$ that is available for establishing an optical link between satellites with 1 W $\textit{P}_T$ at different values of $\textit{d}_{SS}$. Note that $\textit{P}_T$ of Mynaric's laser communication terminal is limited to 1 W \cite{b8}. In this table, when $\textit{d}_{SS}$ is 4,000 km, 4,500 km, 5,000 km, and 5,500 km, the value for $\textit{LM}$ that is available with 1 W $\textit{P}_T$ is 5.6 dB, 4.6 dB, 3.7 dB, and 2.9 dB, respectively, and these results are shown in red in the table. 
\begin{table}
\centering
\renewcommand\thetable{6}
\caption{Transmission Power vs. Link Margin for Optical Inter-Satellite Link.}
\arrayrulecolor{black}
\begin{tabular}{!{\color{black}\vrule}l!{\color{black}\vrule}l!{\color{black}\vrule}l!{\color{black}\vrule}l!{\color{black}\vrule}l!{\color{black}\vrule}} 
\hline
$\textit{d}_{SS}$ (km)                     & \textit{LM} (dB)                      & $\textit{P}_R$ (dBm)      & $\textit{P}_T$ (dBm)                     & $\textit{P}_T$ (W)                           \\ 
\hline
\multirow{5}{*}{~
  ~
  4000}
                                      & 4                                     & -31.5                  & 28.36                                  & 0.686                \\ 
\cline{2-5}
                                      & 5                                     & -30.5                  & 29.36                                   & 0.863                                      \\ 
\cline{2-5}
                                      & \textcolor{red}{5.6}		 & \textcolor{red}{-29.9} 		& \textcolor{red}{29.96  }		 & \textcolor{red}{0.991  }      \\ 
\cline{2-5}
                                      & 6                                     & -29.5                  & 30.36                                  & 1.087                                     \\ 
\cline{2-5}
                                      & 7                                     & -28.5                  & 31.36                                  & 1.369                                     \\ 
\hline
\multirow{5}{*}{~
  ~
  4500}
			 & 3                                     & -32.5                  & 28.39                                   & 0.690                \\ 
\cline{2-5}
                                      & 4                                     & -31.5                  & 29.39                                   & 0.868                                      \\ 
\cline{2-5}
                                      & \textcolor{red}{4.6} 		& \textcolor{red}{-30.9} 		& \textcolor{red}{29.99 }		 & \textcolor{red}{0.997 }     \\ 
\cline{2-5}
                                      & 5                                     & -30.5                  & 30.34                                  & 1.093                                     \\ 
\cline{2-5}
                                      & 6                                     & -29.5                  & 31.39                                  & 1.376                                     \\ 
\hline
\multirow{5}{*}{~
  ~
  5000} 
			& 2                                     & -33.5                  & 28.30                                 & 0.676               \\ 
\cline{2-5}
                                      & 3                                     & -32.5                  & 29.30                                 & 0.851                                     \\ 
\cline{2-5}
                                      & \textcolor{red}{3.7}                  & \textcolor{red}{-31.8}   	& \textcolor{red}{30.00}                & \textcolor{red}{1.000 }  \\ 
\cline{2-5}
                                      & 4                                     & -31.5                  & 30.30                                 & 1.072                                    \\ 
\cline{2-5}
                                      & 5                                     & -30.5                  & 31.30                                  & 1.349                                     \\ 
\hline
\multirow{5}{*}{~
  ~
  5500} 
                                      & 1                                     & -34.5                  & 28.13                                 & 0.650                                     \\ 
\cline{2-5}			
                                      & 2                                     & -33.5                  & 29.13                                 & 0.818                                     \\ 
\cline{2-5}
                                      & \textcolor{red}{2.9}                  & \textcolor{red}{32.6} 	& \textcolor{red}{30.03}                & \textcolor{red}{1.007}                   \\ 
\cline{2-5}
                                      & 3                                     & -32.5                  & 30.13                                 & 1.030                                     \\ 
\cline{2-5}
                                      & 4                                    & -31.5                  & 31.13                                  & 1.297                                     \\ 
\hline
\end{tabular}
\arrayrulecolor{black}
\end{table}

\subsection{Transmission Power vs. Link Margin for Optical Uplink/Downlink}
Table 7 shows that $\textit{P}_T$ increases with increase in $\textit{LM}$ for a certain value of $\textit{d}_{GS}$ when $\textit{$\theta$}_E$ is fixed at 40º. For optical uplink/downlink, we also find $\textit{LM}$ for 1 W $\textit{P}_T$ at different satellite altitudes and thereby different slant distances. In this table, when $\textit{h}_S$ is 600 km, 700 km, 800 km, and 900 km, the $\textit{LM}$ that can be achieved with 1 W $\textit{P}_T$ is 18.3 dB, 17 dB, 15.9 dB, and 15 dB, respectively, and these results are highlighted in red in the table. 

\begin{table}
\centering
\renewcommand\thetable{7}
\setlength{\tabcolsep}{5pt}
\caption{Transmission Power vs. Link Margin for Optical Uplink/Downlink.}
\arrayrulecolor{black}
\begin{tabular}{!{\color{black}\vrule}l!{\color{black}\vrule}l!{\color{black}\vrule}l!{\color{black}\vrule}l!{\color{black}\vrule}l!{\color{black}\vrule}l!{\color{black}\vrule}} 
\hline
$\textit{h}_S$ (km)              & $\textit{d}_{GS}$ (km)                               & \textit{LM} (dB)                       & $\textit{P}_R$ (dBm)      & $\textit{P}_T$ (dBm)   & $\textit{P}_T$ (W)                \\ 
\hline
\multirow{5}{*}{~
  600}                                             & \multirow{5}{*}{~
  881.0}    & 17                    & -18.5                      & 28.70                  & 0.742                   \\ 
\cline{3-6}
                                                                         &                                                                         & 18                                     & -17.5                      & 29.70                & 0.934  \\ 
\cline{3-6}
                                                                         &                                                                         & \textcolor{red}{18.3}                  & \textcolor{red}{-17.2}     & \textcolor{red}{30.00} 	& \textcolor{red}{1.001}             \\ 
\cline{3-6}
                                                                         &                                                                         & 19                                    & -16.5                      & 30.70                & 1.176                            \\ 
\cline{3-6}
                                                                         &                                                                         & 20                                    & -15.5                      & 31.70                & 1.480                            \\ 
\hline
\multirow{5}{*}{~
 700}& \multirow{5}{*}{~
  1020.1} & 15                    & -20.5                      & 27.98                  & 0.628                   \\ 
\cline{3-6}
                                                                         &                                                                         & 16                                     & -19.5                      & 28.98                & 0.790                           \\
\cline{3-6}
                                                                         &                                                                        & \textcolor{red}{17}                  & \textcolor{red}{-18.5}     & \textcolor{red}{29.98} 		& \textcolor{red}{0.995}             \\ 
\cline{3-6}
                                                                         &                                                                         & 18                                     & -17.5                      & 30.98                & 1.252                           \\ 
\cline{3-6}
                                                                         &                                                                         & 19                                    & -16.5                      & 31.98                & 1.577                            \\ 
\hline
\multirow{5}{*}{~
  800}                                         & \multirow{5}{*}{~
  1157.5}                                             & 14                    & -21.5                      & 28.07                  & 0.642                   \\ 
\cline{3-6}
                                                                         &                                                                         & 15                                     & -20.5                      & 29.07                & 0.808                           \\ 
\cline{3-6}
                                                                         &                                                                         & \textcolor{red}{15.9}		 & \textcolor{red}{-19.6}     & \textcolor{red}{29.97}	 & \textcolor{red}{0.994}             \\ 
\cline{3-6}
                                                                         &                                                                         & 16                                     & -19.5                      & 30.07                & 1.017                            \\
\cline{3-6}
                                                                         &                                                                         & 17                                    & -18.5                      & 31.07                & 1.281                            \\  
\hline
\multirow{5}{*}{~
  900}                                         & \multirow{5}{*}{~
  1293.2}                                             & 13                    & -22.5                      & 28.04                  & 0.636                   \\ 
\cline{3-6}
                                                                         &                                                                         & 14                                     & -21.5                      & 29.04                & 0.801                          \\ 
\cline{3-6}
                                                                         &                                                                         & \textcolor{red}{15}                  & \textcolor{red}{-20.5}     & \textcolor{red}{30.04} & \textcolor{red}{1.009}             \\
\cline{3-6}
                                                                         &                                                                         & 16                                     & -19.5                      & 31.04                & 1.270                           \\
\cline{3-6}
                                                                         &                                                                         & 17                                    & -18.5                      & 32.04                & 1.599                            \\ 
\hline
\end{tabular}
\arrayrulecolor{black}
\end{table}

\section{Practical Insights and Design Guidelines}
In this section, we provide important practical insights and design guidelines that can be helpful for practical satellite communication. 

\!\!\!\!\!\!\textbf{Practical Insights}

$\bullet$ For optical inter-satellite link and optical uplink/downlink, $\textit{L}_{PS}$ and $\textit{L}_{PG}$ and thereby $\textit{P}_T$ increase as $\textit{d}_{SS}$ and $\textit{d}_{GS}$ increase. Thereby, the deployment of satellites is of critical importance to maximize their lifetime.

$\bullet$ The $\textit{P}_T$ needed for optical uplink/downlink is larger compared to the one required for optical inter-satellite link with same link distance. For example, for 2,000 km link distance, $\textit{P}_T$ needed for optical inter-satellite link and optical uplink/downlink is 136.20 mW and 152.05 mW, respectively.

$\bullet$ When $\textit{$\theta$}_E$ is fixed at 40º, $\textit{I}_m$ and $\textit{I}_g$ remain the same irrespective of $\textit{h}_S$. This is because both scatterings are independent of $\textit{d}_{GS}$ at a fixed $\textit{$\theta$}_E$ and this indicates that these two attenuations only happen in the atmosphere near the Earth’s surface.

$\bullet$ For optical uplink/downlink, $\textit{I}_m$ and $\textit{I}_g$ and thereby $\textit{L}_A$ vary with $\textit{$\theta$}_E$. However, the relationship between $\textit{$\theta$}_E$ and $\textit{L}_A$ is not linear. $\textit{L}_A$ changes significantly when $\textit{$\theta$}_E$ is small and for large values of $\textit{$\theta$}_E$, the change in $\textit{L}_A$ becomes very small. For example, $\textit{L}_A$ changes significantly when $\textit{$\theta$}_E$ increases from 10º to 20º but changes slightly from 70º to 80º.

$\bullet$ With the increase of $\textit{$\theta$}_E$, $\textit{d}_{GS}$ and $\textit{d}_A$ decrease and reduce $\textit{L}_{PG}$ and $\textit{I}_g$, respectively; $\textit{I}_m$ also decreases with increase in $\textit{$\theta$}_E$ as according to (7) the larger the $\textit{$\theta$}_E$ the lower the $\textit{I}_m$; and this results in a decrease in $\textit{P}_T$ for optical uplink/downlink with increase in $\textit{$\theta$}_E$.

$\bullet$ The $\textit{LM}$ decreases when $\textit{d}_{SS}$ and $\textit{d}_{GS}$ increase at a fixed $\textit{P}_T$, which indicates an inverse relationship. This is because $\textit{L}_{PS}$ and $\textit{L}_{PG}$ increase with increase in $\textit{d}_{SS}$ and $\textit{d}_{GS}$, which degrades $\textit{LM}$ that is available with a fixed $\textit{P}_T$. The $\textit{LM}$ for optical inter-satellite link that is available with 1 W $\textit{P}_T$ decreases when $\textit{d}_{SS}$ increases. A higher $\textit{LM}$ is available at lower $\textit{h}_S$ and thereby lower $\textit{d}_{GS}$ with 1 W $\textit{P}_T$, and the available $\textit{LM}$ decreases with increase in $\textit{h}_S$ and $\textit{d}_{GS}$ for optical uplink/downlink.

\!\!\!\!\!\!\textbf{Design Guidelines}

$\bullet$ With 1 W $\textit{P}_T$ and 3 dB $\textit{LM}$ available to establish a 10 Gbps optical link in an FSOSN, $\textit{d}_{SS}$ is limited to 5,419 km for reliable optical inter-satellite link performance, and $\textit{h}_S$ is restricted to 4,062 km and $\textit{d}_{GS}$ is constrained to 5,125 km for reliable optical uplink/downlink performance. 

$\bullet$ For a 10 Gbps optical inter-satellite link limited by 1 W $\textit{P}_T$, $\textit{LM}$ reaches zero when $\textit{d}_{SS}$ is 7,654 km. The 10 Gbps optical communication link between two satellites can no longer be sustained when $\textit{LM}$ falls below zero as $\textit{P}_R$ falls below $\textit{P}_{req}$. 

$\bullet$ For a 10 Gbps optical uplink/downlink with a limitation of 1 W on $\textit{P}_T$, $\textit{LM}$ reaches zero when $\textit{d}_{GS}$ is 7,240 km and $\textit{h}_S$ is 5,970 km. The satellites in an FSOSN should have $\textit{h}_S$ less than 5,970 km when $\textit{$\theta$}_E$ is 40º and $\textit{P}_T$ is limited to 1 W, as the satellites located at a higher $\textit{h}_S$ will not be able to maintain 10 Gbps optical uplink/downlink communication.

\balance

\section{Conclusion}
In this work, we investigated the link budget for optical inter satellite link and optical uplink/downlink in FSOSNs. We use appropriate system models for these links and study the link transmission power at different link distances, different elevation angles for uplink/downlink, and different link margins. It is observed that with the increase in link distance, the link transmission power increases due to increase in free-space path loss. The results show that the atmospheric attenuation depends upon the elevation angle between the satellite and the ground station for optical uplink/downlink, and this loss increases when the elevation angle decreases. We also investigate the relationship between link margin and link distance for a given link transmission power and link date rate. We observe that the link margin and link distance have an inverse relationship. Furthermore, some practical insights and design guidelines are provided.

In FSOSNs, satellites have an optical inter-satellite link (or laser inter-satellite link (LISL)) range for connectivity. The LISL range is a range within which a satellite can successfully establish an LISL with any other satellite that is within this range. The larger the LISL range, the more the possible connectivity, and the longer the links between satellites. In this way, fewer satellites and LISLs are needed on the path between source and destination ground stations over the FSOSN, which will reduce the latency but will result in an increase in satellite transmission power. In future, we plan to analyze this tradeoff between satellite transmission power and network latency in FSOSNs arising from different LISL ranges. 

\balance

\section*{Acknowledgement}
This work has been supported by the National Research Council Canada’s (NRC) High Throughput Secure Networks program (CSTIP Grant \#CH-HTSN-625) within the Optical Satellite Communications Consortium Canada (OSC) framework.

\end{document}